\pdfoutput=1 
\documentclass[aps,twocolumn,prl,showpacs,amsmath,amssymb,10pt]{revtex4-1}
\usepackage{graphicx}
\usepackage{scrextend}
\usepackage{shuffle}
\usepackage{manfnt,pifont}

\usepackage{url}

\begin{document}


\title{Emergent world-sheet for the AdS Virasoro-Shapiro amplitude}

\author{Luis F. Alday}
\author{Tobias Hansen}
\author{Joao A. Silva}

\affiliation{Mathematical Institute, University of Oxford, Andrew Wiles Building, Radcliffe Observatory Quarter, Woodstock Road, Oxford, OX2 6GG, U.K.}

\date{\today}

\begin{abstract}
\noindent 
We construct a representation for the first $AdS$ curvature correction to the Virasoro-Shapiro amplitude, as an integral over the Riemann sphere. The integrand is that of the Virasoro-Shapiro amplitude in flat space, with the extra insertion of a linear combination of single-valued multiple polylogarithms of weight three. The integral representation implies an elegant, manifestly single-valued representation for the Wilson coefficients of the low energy expansion.

\end{abstract}

\maketitle

\noindent{\bf The idea.} In this letter we study the Virasoro-Shapiro amplitude for the scattering of four gravitons on $AdS_5 \times S^5$. This is defined, via the $AdS/CFT$ correspondence, as the correlator of four stress-tensor multiplets in Mellin space, to leading order in inverse powers of the central charge. The Borel transform of the Mellin amplitude reduces to the usual Virasoro-Shapiro amplitude in the flat space limit, plus a tower of curvature corrections
\begin{equation} 
A(S,T)=A^{(0)}(S,T) + \frac{\alpha'}{R^2} A^{(1)}(S,T) + \cdots
\end{equation}
The Virasoro-Shapiro amplitude in flat space 
\begin{equation}
A^{(0)}(S,T)= -\frac{\Gamma(-S)\Gamma(-T)\Gamma(-U)}{\Gamma(S+1)\Gamma(T+1)\Gamma(U+1)},
\end{equation}
where $S+T+U=0$, admits a low energy expansion 
\begin{equation}
A^{(0)}(S,T)= \frac{1}{S T U} + 2 \sum_{a,b=0}^\infty \sigma_2^a \sigma_3^b \alpha_{a,b}^{(0)},
\end{equation}
with $\sigma_2 = \frac{1}{2}(S^2+T^2+U^2)$ and $\sigma_3= S T U$. It turns out \cite{Stieberger:2013wea} that the Wilson coefficients $\alpha_{a,b}^{(0)}$ live in the ring of single-valued multiple zeta values (MZVs) \cite{Brown:2013gia}. This is manifest in the representation
\begin{equation}
A^{(0)}(S,T)=\frac{ \exp\left( \sum\limits_{n=1}^\infty {\frac{\zeta^{\text{sv}}(2n+1)(S^{2n+1}+T^{2n+1}+U^{2n+1})}{2n+1}} \right) }{S T U},
\end{equation}
where $\zeta^{\text{sv}}(2n+1)=2 \zeta(2n+1)$ are single-valued zeta values. A direct world-sheet computation leads instead to the following representation as an integral over the Riemann sphere
\begin{equation}
A^{(0)}(S,T)=\frac{1}{U^2} \int d^2z  |z|^{-2S-2}|1-z|^{-2T-2},
\label{A0_integral}
\end{equation}
where $z$ denotes the complex cross-ratio on the four-punctured sphere and the integration measure is defined as $d^2 z =dz d \bar z /(-2 \pi i)$. 
It was understood in \cite{Brown:2019wna,Schlotterer:2018zce,Vanhove:2018elu} that the reason for $\alpha_{a,b}^{(0)}$ being in the ring of single-valued MZVs is the single-valued nature \cite{Brown:2018omk} of the integral \eqref{A0_integral}.

The focus of this letter will be the first correction $A^{(1)}(S,T)$. In \cite{Alday:2022uxp,Alday:2022xwz} the low energy expansion for $A^{(1)}(S,T)$ was computed by leveraging Regge boundedness to derive dispersive sum rules and assuming single-valuedness to solve them. The result takes the form

\begin{equation}
A^{(1)}(S,T)= -\frac{2}{3} \frac{ \sigma_2}{ \sigma_3^2} + 2 \sum_{a,b=0}^\infty  \sigma_2^a \sigma_3^b \alpha_{a,b}^{(1)},
\end{equation}
where the Wilson coefficients $\alpha_{a,b}^{(1)}$ live in the ring of single-valued MZVs by construction, justified by the expectation that closed string amplitudes like $A^{(1)}(S,T)$ arise from world-sheet integrals similar to \eqref{A0_integral}. The aim of this letter is to construct such an explicit integral representation for $A^{(1)}(S,T)$, of the form
\begin{equation}
A^{(1)}(S,T)= B^{(1)}(S,T)+ B^{(1)}(U,T)+B^{(1)}(S,U),
\end{equation}
where $ B^{(1)}(S,T)$ is symmetric and given by
\begin{equation}
B^{(1)}(S,T)=\int d^2 z |z|^{-2S-2}|1-z|^{-2T-2} G(z,\bar z).
\end{equation}
Following the discussion in \cite{Vanhove:2018elu} we construct $G(z,\bar z)$ out of single-valued multiple polylogarithms (SVMPLs). This ensures that the low energy expansion of $A^{(1)}(S,T)$ contains only single-valued MZVs. Given that the coincident limit of insertion points in a putative world-sheet correspond to $z \to \{0,1,\infty\}$, we use polylogarithms evaluated at $z$ and labelled by words $w$ with letters $\{0,1\}$. Furthermore, the structure of the Wilson coefficients $\alpha_{a,b}^{(1)}$ found in \cite{Alday:2022uxp,Alday:2022xwz}, together with the order of the poles in $A^{(1)}(S,T)$ (of fourth order and lower) imply that the SVMPLs that appear have at most weight three. We also present a world-sheet motivation for this fact. It turns out the final integral representation is constructed out of SVMPLs of exactly weight three. 

\vspace{0.2cm}
\noindent{\bf Single-valued multiple polylogarithms.} Multiple polylogarithms (MPLs, also known as harmonic polylogarithms) are functions $L_w(z)$  of one variable labelled by a word $w$ formed in our case by the 'letters' 0 and 1. They are recursively defined by the relations
\begin{equation}
\partial_z L_{0w}(z) = \frac{1}{z} L_w(z),~~~~\partial_z L_{1w}(z) = \frac{1}{z-1} L_w(z),
\end{equation}
together with $L_\varnothing(z)=1$, where $\varnothing$ denotes the empty word, and the condition that $\lim_{z \to 0} L_w(z) = 0$ for $w$ not of the form $0^p$. For $w=0^p$ we get $L_{0^p}(z)= \frac{\log^p z}{p!}$. MPLs satisfy shuffle relations 
\begin{equation}
L_{w}(z) L_{w'}(z)  = \sum_{W \in w \shuffle w'} L_{W}(z).
\label{shuffle}
\end{equation}
Evaluated at $z=1$ MPLs reduce to MZVs, for instance
\begin{equation}
L_{ 0^{s_1-1}1 0^{s_2-1} 1 \cdots  0^{s_d-1}1}(1) = \zeta(s_1,\cdots,s_d).
\end{equation}
In the standard notation
\begin{equation}
\zeta(s_1,\cdots,s_d) =\sum_{n_1> \cdots>n_d>0} \frac{1}{n_1^{s_1} \cdots n_d^{s_d}},
\end{equation}
where $s_1+\cdots+s_d$ is the weight, which agrees with the length $|w|$, and $d$ is the depth. When the word starts with $1$, $L_{w}(1)$ is generally divergent. A regularised value can be assigned by using the shuffle relations to isolate the divergent contribution, and then setting $ L_{1}(1)=\zeta(1)=0$ (which is equivalent to setting $\log 0=0$).  As a consequence of (\ref{shuffle}) MZVs also satisfy shuffle relations. MPLs have branch points at $z=0,1,\infty$. A single-valued version of MPLs was constructed in \cite{Brown:2004ugm} via a single-valued map $L_{w}(z) \to {\cal L}_w(z)$ such that $ {\cal L}_w(z)$ is a single-valued weight-preserving linear combination of $L_{w'}(z)L_{w''}(\bar z)$ and satisfies the same differential relations
\begin{equation}
\partial_z {\cal L}_{0w}(z) = \frac{1}{z} {\cal L}_{w}(z),~~\partial_z {\cal L}_{1w}(z) = \frac{1}{z-1} {\cal L}_{w}(z),
\end{equation}
such that ${\cal L}_\varnothing(z)=1$, ${\cal L}_{0^p}(z)= \frac{\log^p |z|^2}{p!}$ and $\lim_{z\to 0} {\cal L}_w(z) = 0$ for $w$ not of the form $0^p$. Furthermore they also satisfy the same shuffle relations
\begin{equation}
{\cal L}_{w}(z) {\cal L}_{w'}(z)  = \sum_{W \in w \shuffle w'} {\cal L}_{W}(z).
\end{equation}
Evaluated at $z=1$ they give rise to single-valued MZVs. For instance 
\begin{equation}
{\cal L}_{ 0^{k_1-1}1  0^{k_2-1} 1 \cdots 0^{k_d-1}1}(1) = 
\zeta^{\text{sv}}(k_1,\cdots,k_d).
\end{equation}
Single-valued MZVs are a subset of the standard MZVs and satisfy several relations, including the same shuffle relations. Furthermore, one can check
\begin{equation}
\zeta^{\text{sv}}(2n) = 0,~~~\zeta^{\text{sv}}(2n+1)= 2\zeta(2n+1) .
\end{equation}
For higher depths the single-valued map acts in a complicated way, but can be computed for instance with the program HyperlogProcedures \cite{HyperlogProcedures}.

\vspace{0.2cm}
\noindent{\bf Integrating SVMPLs.} Next we consider the integrals
\begin{equation}
I_w(S,T)= \int d^2z |z|^{-2S-2}|1-z|^{-2T-2} {\cal L}_w(z),
\end{equation}
and compute their low energy expansion. \cite{Vanhove:2018elu} have considered the low energy expansion for the particular case ${\cal L}_\varnothing(z)=1$ in great detail, and developed a machinery to treat the general case. In particular we will use the following two key results. First, given any SVMPL we can write
\begin{equation}
\frac{{\cal L}_w(z)}{|z|^2|1-z|^2} =\partial_z F_w(z) ,
\end{equation}
where $F_w(z)$ is given by
\begin{equation}
F_w(z) = \frac{{\cal L}_{0w}(z)-{\cal L}_{1w}(z)}{\bar z (1-\bar z)}.
\end{equation}
This follows from the differential relations that SVMPLs satisfy. The second result regards the integration of SVMPLs. For the case at hand
\begin{equation}
 \int d^2z \partial_z F(z) = \overline{Res}_{z=\infty} F(z) - \overline{Res}_{0} F(z)-\overline{Res}_{1} F(z) .
\end{equation}
This follows from Stokes theorem \cite{Schnetz:2013hqa}. Here the residues around infinity, zero and one are defined by the expansions around these points. For SVMPLs we have
\begin{equation}
{\cal L}_{w}(z) = \sum_{m,n,k}  c_{kmn} (z-\sigma)^m(\bar z - \bar \sigma)^n \log^k|z-\sigma|^2.
\end{equation}
The anti-holomorphic residue around a point $z=\sigma$ is defined as the coefficient $c_{0,0,-1}$.


Let us now turn to the problem at hand.  Following \cite{Vanhove:2018elu} we split the integral above into two contributions
\begin{align}
I^{(1)}_w(S,T)&= \int d^2z \frac{(|z|^{-2S}-1)(|1-z|^{-2T}-1)}{|z|^2|1-z|^2} {\cal L}_w(z), \nonumber\\
I^{(2)}_w(S,T)&= \int d^2z \frac{|z|^{-2S}+|1-z|^{-2T}-1}{|z|^2|1-z|^2}  {\cal L}_w(z).
\end{align}
Let us consider the first contribution. This is absolutely convergent at $S=T=0$, and hence we can Taylor expand around that point, and exchange summation and integration. This leads to 
\begin{equation}
I^{(1)}_w(S,T)=\sum_{p,q=1} (-S)^p (-T)^q  \int d^2z  \frac{{\cal L}_{0^p}(z) { \cal L}_{1^q}(z) {\cal L}_w(z)}{|z|^2|1-z|^2}.
\end{equation}
Using the shuffle product we write 
\begin{equation}
  {\cal L}_{0^p}(z) { \cal L}_{1^q}(z) {\cal L}_w(z)= \sum_{W \in 0^p \shuffle 1^q \shuffle w}  {\cal L}_W(z).
\end{equation}
Then, using the two key results above we obtain
\begin{equation}
I^{(1)}_w(S,T)= \sum_{p,q=1}  (-S)^p (-T)^q  \hspace{-13pt} \sum_{W \in 0^p \shuffle 1^q \shuffle w}  \hspace{-13pt} \left( {\cal L}_{0W}(1)-{\cal L}_{1W}(1) \right).
\end{equation}
This generalises in an obvious way the result by \cite{Vanhove:2018elu} valid for $w=\varnothing$. Let us now turn to the second contribution. Following again \cite{Vanhove:2018elu} we consider
\begin{equation}
I^{(2),\epsilon}_w(S,T)= \sum_{ \substack{p,q \geq 0 \\ p.q=0}}  (-S)^p (-T)^q \sum_{W} \int_{U_\epsilon} \partial_z F_W(z) d^2z,
\end{equation}
where $W \in 0^p \shuffle 1^q \shuffle w$ and $U_\epsilon = \mathbb{C} \setminus (B_0(\epsilon) \cup B_1(\epsilon) \cup B_0(\epsilon^{-1})) $ is the complex plane minus three balls, around zero, one and infinity. Clearly $ I^{(2)}_w(S,T)=\lim_{\epsilon \to 0}  I^{(2),\epsilon}_w(S,T)$. By Stokes theorem this receives only contributions from the three boundaries
\begin{align}
I^{(2)}_w(S,T) ={}&\sum \limits_{ \substack{p,q \geq 0 \\ p.q=0}}(-S)^p (-T)^q \sum \limits_{W}  \bigg( \hspace{-8pt} \int \limits_{\partial^+ B_0(\epsilon^{-1})} \hspace{-8pt} F_W(z)  \frac{i d\bar z}{2\pi} \nonumber\\
&+ \hspace{-8pt}  \int \limits_{\partial^- B_0(\epsilon)} \hspace{-8pt} F_W(z)  \frac{i d\bar z}{2\pi} 
+ \hspace{-8pt} \int \limits_{\partial^- B_1(\epsilon)} \hspace{-8pt} F_W(z)  \frac{i d\bar z}{2\pi}\bigg),
\end{align}
with $F_W(z) = \frac{{\cal L}_{0W}(z)-{\cal L}_{1W}(z)}{\bar z (1-\bar z)}$. Each boundary can be analysed separately. Let us assume for simplicity $w$ is not of the form $0^n$, and return to this case later. In this case the contribution from the boundary at zero vanishes, and the same is true for the boundary at infinity. The contribution from $\partial^- B_1(\epsilon)$ is slightly more subtle and in order to compute it we need the behaviour of ${\cal L}_{w}(z)$ around $z=1$. For $w=1^n$ we have ${\cal L}_{1^n}(1+\epsilon) = \frac{\log^n |\epsilon|^2}{n!}$. For $w$ starting with a $0$ we have
\begin{equation}
{\cal L}_{0w'}(1+\epsilon) = {\cal L}_{0w'}(1) + {\cal O}(\epsilon) ,
\end{equation}
for finite $ {\cal L}_{0w'}(1) $ given in terms of single-valued MZVs. When the word starts with $1$ we can use the shuffle identities to isolate and compute its logarithmic divergences as we approach $z=1$. Assume for instance $w=10w'$, then
\begin{equation}
{\cal L}_{1}(z) {\cal L}_{0w'}(z) = {\cal L}_{w}(z) +  \sum_{W \in 1 \shuffle w'} {\cal L}_{0W}(z) ,
\end{equation}
which leads to 
\begin{equation}
 {\cal L}_{10w'}(1+\epsilon) = \log |\epsilon|^2 {\cal L}_{0w'}(1)  - \sum_{W \in 1 \shuffle w'} {\cal L}_{0W}(1) + {\cal O}(\epsilon) .
\end{equation}
Applying this idea recursively, we can compute the logarithmic behaviour, around $z=1$, for any word. The contribution from the boundary at $1$ will be of the form 
\small{
\begin{equation}
I^{(2)}_w(S,T)= \text{polar}+\sum_{ \substack{p,q \geq 0 \\ p.q=0}}(-S)^p (-T)^q   \hspace{-15pt} \sum_{W\in 0^p \shuffle 1^q \shuffle w} \hspace{-15pt} \left({\cal L}_{0W}(1)-{\cal L}_{1W}(1) \right).
\end{equation}
}
Combining both contributions we obtain
\small{
\begin{equation}
I_w(S,T)= \text{polar}+\sum_{p,q=0} (-S)^p (-T)^q \hspace{-15pt} \sum_{W\in 0^p \shuffle 1^q \shuffle w} \hspace{-15pt} \left( {\cal L}_{0W}(1)-{\cal L}_{1W}(1) \right).
\end{equation}
}
This result is valid for arbitrary $w$. The polar contribution arises from logarithmic divergences either around $z=0$, for $w=0^n$ or around $z=1$, as explained above. For example
\begin{equation}
\text{polar}(0^n)= -\frac{1}{S^{n+1}},~~~\text{polar}(1^n)= -\frac{1}{T^{n+1}}. 
\end{equation}
In general, a logarithmic divergence $ {\cal L}_{1^q}(1+\epsilon)$ leads to a polar term $-1/{T^{q+1}}$. Below we will need $I_w(S,T)$ for weight three. An example of this is
\begin{equation}
I_{101}(S,T)= \frac{4 \zeta(3) }{T} +2 (2T-S) \zeta(5) +2(4T^2+4 S T-S^2) \zeta(3)^2 + \cdots
 \end{equation}
In particular we see the generic term is an homogeneous polynomial of degree $n$ times a (single-valued) MZV of weight $n+4$. This is true for all other cases of weight three, as well as for $\zeta(3) I_\varnothing(S,T)$. 

\vspace{0.2cm}
\noindent{\bf Integral representation for $A^{(1)}(S,T)$.} The coefficients in the low energy expansion of $I_w(S,T)$ are in the ring of single-valued MZVs. Hence, we will attempt to construct an integral representation for $A^{(1)}(S,T)$ such that 
\begin{equation}
A^{(1)}(S,T)= B^{(1)}(S,T)+ B^{(1)}(U,T)+B^{(1)}(S,U),
\end{equation}
where $ B^{(1)}(S,T)$ is symmetric and of the form
\begin{equation}
B^{(1)}(S,T) = \sum_w R_w(S,T) I_w(S,T),
\end{equation}
for some rational functions $R_w(S,T)$. From the low energy expansion for $A^{(1)}(S,T)$
\begin{equation}
A^{(1)}(S,T) = -\frac{2}{3} \frac{\sigma_2}{\sigma_3^2} - \frac{44}{3} \zeta(3)^2  \sigma_2 - \frac{537}{8} \zeta(7) \sigma_3 + \cdots
\end{equation}
we see that the generic term is a homogeneous polynomial of degree $n$ times MZVs of weight $n+4$. This suggests $w$ in our ansatz should have weight three and the rational functions be the ratio of two homogeneous polynomials of the same degree. This is confirmed by the pole structure of $A^{(1)}(S,T)$ with quartic (and lower) order poles. Indeed, note that the insertion of $\log^3 |z|^2$ is equivalent to taking $\partial_S^3$, which increases the degree of the single poles of $A^{(0)}(S,T)$ by three.

Next, note that due to the $z \leftrightarrow \bar z$ symmetry of the integrand, the integral $I_w(S,T)$ is only sensitive to the symmetric part of  ${\cal L}_w(z)$. Using the explicit expressions for ${\cal L}_w(z)$ for weight three, given in the appendix, this implies 
\begin{equation}
I_{011}(S,T) =I_{110}(S,T),~~~I_{001}(S,T) =I_{100}(S,T).
\end{equation}
This reduces the number of independent integrals $I_w(S,T)$ at weight three from eight to six. Next, we want to construct a function $B^{(1)}(S,T) $ which is symmetric under $S \leftrightarrow T$. By a change of variables $z \to 1-z$ we see 
\begin{equation}
I_w(T,S)= \int d^2z |z|^{-2S-2}|1-z|^{-2T-2} {\cal L}_w(1-z).
\end{equation}
SVMPLs are closed under this transformation. For weight three
\begin{equation}
\begin{aligned}
{\cal L}_{000}(1-z) &= {\cal L}_{111}(z), \\
{\cal L}_{001}(1-z) &= {\cal L}_{110}(z) -2 \zeta(3),\\
{\cal L}_{010}(1-z) &= {\cal L}_{101}(z) +4 \zeta(3),\\
{\cal L}_{100}(1-z) &= {\cal L}_{011}(z) -2 \zeta(3),
\end{aligned}
\end{equation}
so that for instance $I_{001}(T,S)=I_{110}(S,T) - 2 \zeta(3) I_\varnothing(S,T)$ and so on. Note furthermore that $I_\varnothing(S,T)$ is itself symmetric. We then try the following ansatz
\begin{align}
B^{(1)}(S,T) ={}& R_0(S,T) I_{000}(S,T)+R_0(T,S) I_{111}(S,T) \label{ansatz} \\
&+R_1(S,T)I_{010}(S,T) +R_1(T,S)I_{101}(S,T) \nonumber\\
&+R_2(S,T)I_{001}(S,T)+R_2(T,S) I_{110}(S,T) \nonumber\\
&+  \left(R_{\text{asym}}(S,T)+ R_{\text{sym}}(S,T)\right) \zeta(3) I_\varnothing(S,T), \nonumber
\end{align}
where based on the integral representation for $A^{(0)}(S,T)$ we propose rational functions of the form $R_i(S,T) = U^{-2} P^{(2)}_i(S,T)$, for $P^{(2)}_i(S,T)$ homogeneous polynomials of degree 2. In the above ansatz $R_{\text{asym}}(S,T) =2(R_1(T,S)-R_1(S,T))-R_2(T,S)+R_2(S,T)$ is an anti-symmetric function fixed so that $B^{(1)}(S,T)$ is symmetric. Finally $R_{\text{sym}}(S,T)$ cannot be fixed by our procedure, and we set it to zero. In total our ansatz contains only 9 coefficients, which are fixed by matching the low energy expansion with $A^{(1)}(S,T)$. We find the following remarkably simple solution
\begin{equation}
\begin{aligned}
R_0(S,T) &= -\frac{2 S}{3 (S+T)},&R_1(S,T)&=-\frac{S+5 T}{6 (S+T)},\\
R_2(S,T) &=\frac{2 (S-T)}{3 (S+T)},&R_{\text{asym}}(S,T)&=0.
\end{aligned}
\end{equation}

\vspace{0.2cm}
\noindent{\bf Structure of poles.} $A^{(1)}(S,T)$ has a very interesting structure of poles that we would like to reproduce from its integral representation. Let us write 

\begin{equation}
B^{(1)}(S,T)=\int d^2 z |z|^{-2S-2}|1-z|^{-2T-2} G(S,T,z),
\end{equation}
where for the discussion below it is important to include the explicit dependence of the integrand on $S,T$. By a change of variables it is easy to see that
\begin{align}
B^{(1)}(U,T)&=\int d^2 z |z|^{-2S-2}|1-z|^{-2T-2} |z|^2 G(U,T,1/z),\nonumber\\
B^{(1)}(S,U)&=\int d^2 z |z|^{-2S-2}|1-z|^{-2T-2} |1-z|^2 G(S,U,\tfrac{z}{z-1}).
\end{align}
The full expression is then 
\begin{equation}
A^{(1)}(S,T)=\int d^2 z |z|^{-2S-2}|1-z|^{-2T-2} G_{\text{tot}}(S,T,z),
\end{equation}
with 
\begin{equation}
\begin{aligned}
G_{\text{tot}}(S,T,z) ={}& G(S,T,z)+|z|^2 G(U,T,1/z) \\
&+|1-z|^2 G(S,U,\tfrac{z}{z-1}).
\end{aligned}
\end{equation}
Note that single-valued polylogarithms are closed under $z \to 1/z$ and $z \to \frac{z}{z-1}$ so the above expression can again be expressed in terms of ${\cal L}_w(z)$. We are interested in computing the poles at $S=0,1,\dots$. These arise from the region of integration around $z=0$ and can be computed by expanding around this point and using polar coordinates. Around $z=0$
\begin{equation}
|1-z|^{-2T-2} G_{\text{tot}}(S,T,z)  = \frac{S^2}{9 T(S+T)} \log^3 |z|^2+ \cdots \label{polarexpansion}
\end{equation}
The leading order term leads to poles at $S=0$
\begin{eqnarray}
\left. A^{(1)}(S,T) \right|_{S=0} &=& \int_0^{\rho_0} \rho d\rho \int_0^{2 \pi} \frac{d\phi}{\pi}  \rho^{-2 S-2} \frac{S^2}{9 T(S+T)} \log^3 \rho^2  \nonumber\\
&=&-\frac{2}{3 S^2 T^2}+ \frac{2}{3 S T^3} + \text{regular},
\end{eqnarray}
which is the correct behaviour for $A^{(1)}(S,T)$ around $S=0$. Keeping higher order terms in the expansion (\ref{polarexpansion}) we can compute poles at $S=1,2,\cdots$, for instance, to next order we obtain
\begin{equation}
A^{(1)}(S,T) =\frac{-1 }{(S-1)^4}+\frac{2/3  }{(S-1)^3}+\frac{2/3  }{(S-1)^2}-\frac{2 \zeta(3)-1}{S-1}+ \text{reg.}
\end{equation}
which again agrees with the expected terms.  The following comment is in order. From the computation in \cite{Alday:2022uxp,Alday:2022xwz} it turns out that quartic and cubic poles follow from the spectrum at leading order, which is the flat space string spectrum. As a result, quartic and cubic poles have a very simple form. It turns out however, that matching these poles is not enough to fix the form of our answer.
Instead, some of the terms in the ansatz \eqref{ansatz} depend on the $AdS$ corrections to the OPE data, which was determined in \cite{Alday:2022xwz}.

\vspace{0.2cm}
\noindent{\bf A world-sheet perspective.}  Although a direct world-sheet computation for string amplitudes on $AdS_5 \times S^5$ is out of reach at the moment, we can understand the main feature of our result from the following world-sheet model. Consider the first $AdS$ curvature correction to the flat-space tree-level four-graviton amplitude. Around flat space the $AdS$ metric takes the form $g_{\mu \nu}(X)= \eta_{\mu \nu} + \frac{h_{\mu \nu}}{R^2} + \cdots$ with $h_{\mu \nu} \sim X_\mu X_\nu$. Hence, one of the contributions to such a curvature correction is given by the integrated insertion of an extra graviton vertex operator with polarisation vector $h_{\mu \nu}$ 
\begin{equation}
h_{\mu \nu} \sim X_\mu X_\nu \sim   \lim_{P_g \to 0}  \frac{\partial^2 }{\partial_{P_g^\mu}\partial_{P_g^\nu}}  e^{i P_g \cdot X},
\end{equation}
where $P_g$ is the momentum of the inserted graviton. We are then led to compute
\begin{equation}
\lim_{P_g \to 0}  \frac{\partial^2 }{\partial_{P_g^\mu}\partial_{P_g^\nu}}  \int d^2 u   \langle V_1(0) V_2(1) V_3(\infty) V_4(z) V_g^{\mu \nu}(u) \rangle.
\end{equation}
The superstring amplitude for the scattering of five gravitons can be written in terms of a set of building blocks, see \cite{Schlotterer:2012ny}. A prototype of the integrals to compute is
\begin{equation}
\lim_{P_g \to 0} \frac{\partial^2 }{\partial_{P_g^\mu}\partial_{P_g^\nu}}   {\cal J}(z,\bar z),
\end{equation}
where
\begin{equation}
 {\cal J}(z,\bar z)=\int d^2 u  \frac{|u|^{2a} |1-u|^{2b}|u-z|^{2c}}{|u|^2|1-u|^2}.
\end{equation}
The aim is to compute $ {\cal J}(z,\bar z)$ as an expansion for small $(a,b,c)=\alpha'(P_1 \cdot P_g,P_2 \cdot P_g,P_4 \cdot P_g)$ to quadratic order. In \cite{Vanhove:2020qtt} this integral was written in terms of products of holomorphic/anti-holomorphic integrals
\begin{equation}
{\cal J}(z,\bar z)=-\frac{1}{\pi} \left( \kappa_1 J_1(z)J_1(\bar z)+\kappa_2 J_2(z)J_2(\bar z) \right),
\end{equation}
where $\kappa_1=\sin (\pi  a) \csc (\pi  (b+c)) \sin (\pi  (a+b+c))$, $\kappa_2=\sin (\pi  b) \sin (\pi  c) \csc (\pi  (b+c))$ and 
\begin{equation}
\begin{aligned}
J_1(z) &= \int_{-\infty}^0(-u)^{a-1}(1-u)^{b-1}(z-u)^cdu,\\
J_2(z) &= \int_{z}^1 u^{a-1}(1-u)^{b-1}(u-z)^c du.
\end{aligned}
\end{equation}
These integrals can be solved in terms of hypergeometric functions and expanded for small $a,b,c$
\begin{equation}
\begin{aligned}
J_1(z) &= \frac{1}{a} + \frac{c}{a} \log z+\cdots,\\
J_2(z) &= \frac{1}{b}+\frac{b+c}{c} \log(1-z)+\cdots.
\end{aligned}
\end{equation}
Expanding ${\cal J}(z,\bar z)$ for small $a,b,c$ we obtain
\begin{align}
{\cal J}(z,\bar z) &= -\frac{1}{a} - \frac{1}{b} -\frac{c}{a} {\cal L}_0(z)-\frac{c}{b} {\cal L}_1(z) \\
& +c \left( {\cal L}_0(z){\cal L}_1(z)-\frac{a+c}{2a} {\cal L}^2_{0}(z)-\frac{b+c}{2b} {\cal L}^2_{1}(z) \right)+\cdots,
\nonumber
\end{align}
so that at $n-th$ order in $a,b,c$ we obtain SVMPLs of weight $n+1$. This implies
\begin{equation}
\lim_{P_g \to 0} \frac{\partial^2 }{\partial_{P_g^\mu}\partial_{P_g^\nu}}   {\cal J}(z,\bar z) = \text{div.} +d \, \zeta_3+ \sum_{i,j,k=0,1} d_{ijk}  {\cal L}_{ijk}(z).
\end{equation}
One shouldn't worry about the divergent contribution, since this is only a small part of the computation. At finite order we get a combination of SVMPLs of weight three, exactly as in our ansatz. We expect other contributions to behave in the same way. 

\vspace{0.2cm}
\noindent{\bf Discussion.} In this letter we have given an integral representation for the first $AdS$ curvature correction to the Virasoro-Shapiro amplitude in flat space. It takes the form of the genus zero world-sheet integral for the usual Virasoro-Shapiro amplitude in flat space with the extra insertion of single-valued multiple polylogarithms of weight three. This feature can be understood from a simple world-sheet model.  Going to higher orders, it would be interesting to understand if a similar representation still holds, and what is the relevant space of single-valued functions. In the simplest scenario, at the next order SVMPLs of weight six would suffice. A direct world-sheet computation from string theory on $AdS_5 \times S^5$ is out of reach at the moment. However, over the last years there has been promising progress in the construction of vertex operators in the pure spinor formalism \cite{Berkovits:2019rwq,Fleury:2021ieo}. Note that our final result is remarkably simple, and it should serve as a guiding target towards defining and computing graviton amplitudes in the pure spinor formalism. 

\vspace{0.2cm}
\noindent{\bf Acknowledgements.} We thank Francis Brown, Oliver Schnetz and Federico Zerbini for useful discussions.
The work of LFA and TH is supported by the European Research Council (ERC) under the European Union's Horizon
2020 research and innovation programme (grant agreement No 787185). LFA is also supported in part by the STFC grant ST/T000864/1. JS is supported by the STFC grant ST/T000864/1. 

\appendix

\vspace{0.2cm}
\noindent{\bf Appendix: Single-valued multiple polylogarithms.}
Below we give the single-valued multiple polylogarithms of weight three in an explicit form. We build them from multiple polylogarithms which up to weight three are given by

\begin{align}
L_{01}(z) ={}& -\text{Li}_2(z),~~~~L_{10}(z) =\text{Li}_2(z)+\log (1-z) \log (z), \nonumber\\
L_{001}(z) ={}& -\text{Li}_3(z),~~~L_{010}(z) =2 \text{Li}_3(z)-\text{Li}_2(z) \log (z),\nonumber\\
L_{100}(z) ={}&-\text{Li}_3(z)+\text{Li}_2(z) \log (z)+\tfrac{1}{2} \log (1-z) \log ^2(z),\nonumber\\
L_{110}(z) ={}& -\text{Li}_3(1-z)+\tfrac{1}{6} \pi ^2 \log (1-z)+\zeta (3),\nonumber\\
L_{101}(z) ={}& 2 \text{Li}_3(1-z)-2 \zeta (3) 
-\log (1-z) \big(2 \text{Li}_2(1-z)\nonumber\\
& +\text{Li}_2(z)+\log (1-z) \log (z)\big),\nonumber\\
L_{011}(z) ={}&-\text{Li}_3(1-z)+\text{Li}_2(1-z) \log (1-z),\\
& +\tfrac{1}{2} \log (z) \log ^2(1-z)+\zeta (3),\nonumber
\end{align}
together with $L_{0^p}(z)=\frac{\log^p z}{p!}$, $L_{1^p}(z)=\frac{\log^p(1-z)}{p!}$. Single-valued multiple polylogarithms are constructed from those. In particular,  see {\it e.g.}\ \cite{Dixon:2012yy}
\begin{equation}
\begin{aligned}
{\cal L}_{000}(z) &= L_{000}(z)+ L_{000}(\bar z)+ L_{00}(z)L_{0}(\bar z)+ L_{0}(z)L_{00}(\bar z),\\
{\cal L}_{001}(z) &= L_{001}(z)+ L_{100}(\bar z)+ L_{00}(z)L_{1}(\bar z)+ L_{0}(z)L_{10}(\bar z),\\
{\cal L}_{010}(z) &= L_{010}(z)+ L_{010}(\bar z)+ L_{01}(z)L_{0}(\bar z)+ L_{0}(z)L_{01}(\bar z),\\
{\cal L}_{100}(z) &= L_{100}(z)+ L_{001}(\bar z)+ L_{10}(z)L_{0}(\bar z)+ L_{1}(z)L_{00}(\bar z),\\
{\cal L}_{110}(z) &= L_{110}(z)+ L_{011}(\bar z)+ L_{11}(z)L_{0}(\bar z)+ L_{1}(z)L_{01}(\bar z),\\
{\cal L}_{101}(z) &= L_{101}(z)+ L_{101}(\bar z)+ L_{10}(z)L_{1}(\bar z)+ L_{1}(z)L_{10}(\bar z),\\
{\cal L}_{011}(z) &= L_{011}(z)+ L_{110}(\bar z)+ L_{01}(z)L_{1}(\bar z)+ L_{0}(z)L_{11}(\bar z),\\
{\cal L}_{111}(z) &= L_{111}(z)+ L_{111}(\bar z)+ L_{11}(z)L_{1}(\bar z)+ L_{1}(z)L_{11}(\bar z).
\end{aligned}
\end{equation}

\bibliography{refletter}
\bibliographystyle{utphys}

\end{document}